\documentclass [10pt,a4paper]{article}
\usepackage[margin=2.5cm]{geometry}
\usepackage{graphicx} 
\usepackage{xspace}
\usepackage{url}

\title{Money: Who Has a Stake in the Most Value-Centric Common Design Material?}
\author{Ryan Bowler, Chris Speed, Geoffrey Goodell}
\date{May 2023}

\begin{document}
\maketitle


\section{Abstract}

Money is more than just a numeric value.  It embodies trust and moral gravity, and it offers flexible ways to transact. However, the emergence of Central Bank Digital Currency (CBDC) is set to bring about a drastic change in the future of money. This paper invites designers to reflect on their role in shaping material and immaterial monetary change. In this rapidly changing landscape, design could be instrumental in uncovering and showcasing the diverse values that money holds for different stakeholders. Understanding these diversities could promote a more equitable and inclusive financial, social, and global landscape within emergent forms of cash-like digital currency. Without such consideration, certain forms of money we have come to know could disappear, along with the values people hold upon them. We report on semi-structured interviews with stakeholders who have current knowledge or involvement in the emerging field of Central Bank Digital Currency (CBDC). Our research indicates that this new form of money presents both challenges and opportunities for designers. Specifically, we emphasise the potential for Central Bank Digital Currency (CBDC) to either positively or negatively reform values through its design. By considering time, reflecting present values, and promoting inclusion in its deployment, we can strive to ensure that Central Bank Digital Currency (CBDC) represents the diverse needs and perspectives of its users.

\section{Introduction}

Throughout history, currency has taken many forms, from shells and metal coins to paper money and, more recently, non-regulated cryptocurrencies. As governments explore their own versions of digital currency, such as Central Bank Digital Currencies (CBDC), design has an increasingly important role to play in preserving the many values associated with money. This paper invites designers to reflect on their role in shaping or maintaining the values associated with money, both established forms like cash and new forms like Central Bank Digital Currency.

Money represents more than just a numeric value to people: it embodies trust, serves a social role, and offers flexible ways to transact through its tangible and intangible properties. In doing so, money binds social relationships~\cite{halloluwa2018value} and can be invaluable to certain demographic groups in their daily routines~\cite{vines2012eighty}. Money underpins community values. Equally, it can produce new forms of value through its usage~\cite{speed2022future}. Money binds many parts of society together~\cite{ingham1996money}. If the diversity of value and use that money has come to represent, both materially and immaterially, diminishes, as cash has already begun to do~\cite{UKDigitalPound2023}, then there is a real risk that the people who rely on these forms of money will become excluded and that the realisation of their values will be impacted.

As we look towards the fast-approaching horizon of CBDC, our findings suggest that ongoing discussion proceeds at a breathless pace while leaving many questions unanswered.  We report that CBDC can offer new ways to create value, for example, in environmental sustainability, but it cannot ignore existing features upon which users of money depend, for example, the anonymity of cash that allows its users to avoid discrimination.  Furthermore, understanding the design of CBDC can allow us to evaluate which user perspectives have been included, and which have been excluded, in shaping its design. We evaluate these considerations, using semi-structured interviews with stakeholders who have current knowledge or involvement in the field. Much of what is known to date on this digital currency is theoretical, speculative, or technical and its social, political, economic, and global impact is yet to be realised. This paper serves as a starting point, enabling designers to consider their role in shaping or reinforcing the values associated with money, both established forms like cash and new forms like CBDC.

\subsection{The Materiality and Social Interconnections of Money}

Money can be thought of from an engineering perspective. It is a material or fabric that is transformed into a product~\cite{sousa2021material}. Traditionally, it significantly comprises metal or paperlike material that is crafted into a tangible asset that can be exchanged or swapped~\cite{ingham2013nature} 
Some have claimed that money has been essential since the beginning of humanity in various forms, from shells to coins, as it represents trust and the promise to repay or purchase~\cite{arvidsson2019building}. Underpinning these materials is debt, classed as a fundamental element that ignites the essential design principles of money:

\begin{quote}

\textit{``Moneys — both “near” and proper — exhibit three principal design components: they are made of debt; that debt is specifically fashioned to create liquidity; and the debt (or credit) medium that results comes with a pledge of value. Those design components recur across major money types, including medieval coins, modern base (high-powered) money, and commercial bank money.}''~\cite{Desan2021}

\end{quote}

The concept of debt significantly shapes the materials of money, resulting in the creation of assets such as cash, stocks, bonds, and real estate. Each of these assets has its own value and methods for conversion into cash. It can be easy to equate the design of money with the design of tangible materials, focussing on their physical affordances and limitations. However, with digitisation, the nature of money is changing. 

In practice, money is increasingly becoming an immaterial store of value in digital form~\cite{peneder2022digitization}. This means that physical forms of money, such as coins and banknotes, are being replaced by digital representations that can be stored and transferred electronically. As money becomes increasingly digital, economists race to question what money is in its digital form, while businesses continue, at pace, to digitise the financial landscape~\cite{peneder2022digitization}. 

Digital representations of money are popular in part thanks to a wider technological industry known as financial technology, or ``FinTech''. FinTech comprises mobile banking, financial apps, online stock trading, cryptocurrencies, and various other mechanisms that support financial transactions via telecommunications networks. FinTech offers financial companies the capability of using digital money and offers non-financial companies the ability to integrate financial services into their products~\cite{worldbank2023}.  The introduction of buy-now-pay-later schemes by Amazon and the introduction of credit cards by Apple are two examples.  FinTech has produced financial innovations that have increasingly had a material impact on the financial markets, institutions, and services with which people interact daily~\cite{schindler2017fintech}. As consumers enjoy the convenience of tapping their phones to make payments, FinTech gathers valuable data from their actions.  As financial transactions become linked with digital identity, they are transformed into valuable assets~\cite{speed2022future}, with important consequences for privacy and human rights~\cite{armer1968,armer1975,nissenbaum2017}.  These technologies can provide insights into social activities such as shopping by offering services like buy now pay later schemes or subscription services, which have seen an increase in popularity.



The digitisation of money through technology requires careful consideration. Money plays a crucial role in social, economic, and political discourses~\cite{ingham2020money}, and its digital forms must be critically evaluated to understand their potential impact on societies and individual values. One way to apply a critical lens to FinTech is by assessing its claims, such as its ability to mitigate issues like financial marginalisation. Inclusion is contextualised through the use of technology to create services that facilitate access to financial services for individuals who lack relationships with banks.~\cite{gopane2019enquiry}. However, not everyone has access to the necessary technology, infrastructure, education, and trust to benefit from these inclusion efforts. This is evident when commercial banks acknowledge that 11 million individuals in the UK cannot use digital financial services~\cite{lloyds2022consumer}. By carefully considering the potential benefits and drawbacks of new FinTech solutions~\cite{drs2022}, design should not only address the technological aspects of digital money but also the social barriers that impede access to an equitable financial landscape. This should be accessible whether money is tangible or digital, as it should not be assumed that individuals want to engage with digital forms of money, even if they have the ability to do so. This responsibility is not only the onus of designers but requires collaboration from stakeholders across society, to ensure that protections are in place to prevent exploitation, particularly for those who may already be disadvantaged by social inequalities.~\cite{ozili2020theories}. Moreover, as FinTech allows finance to bridge international boundaries~\cite{worldbank2023}, the voices of global stakeholders are no less important.

\begin{figure}[h]
   \centering
    \scalebox{0.9}{\includegraphics[width=0.5\textwidth]{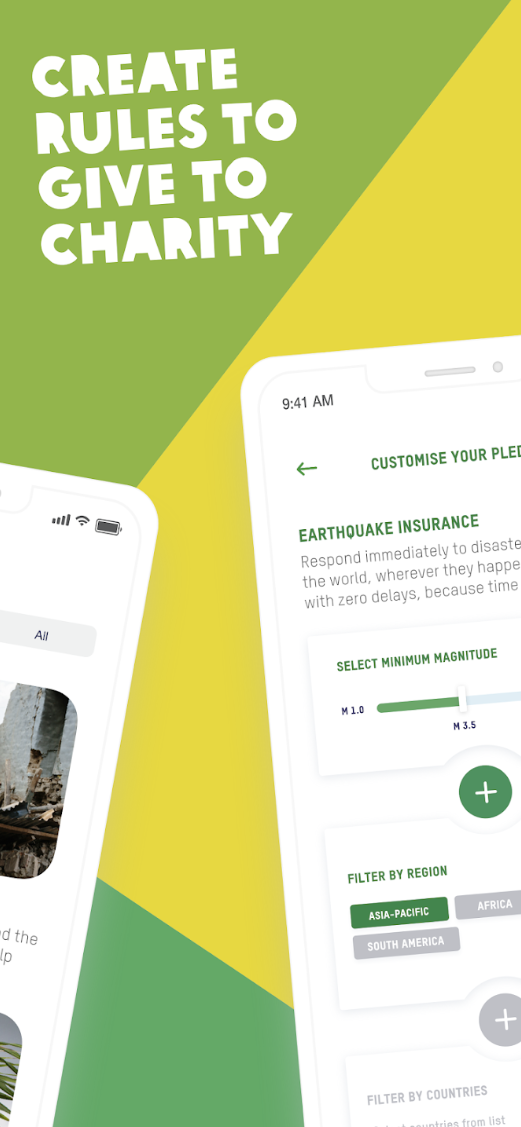}}
    \caption{Oxfam: If Then Give.}
   \label{fig:Oxfam: If Then Give}
\end{figure}

Design research has already begun to demonstrate that technologies like smart contracts can be used to envision new ways of transacting money that has with them social and global benefits.
Some methods of doing this involve leveraging the communicative properties of social media platforms, such as use of `\texttt{\#}' and `\verb+@+' symbols to refer to threads of conversations and individual users, respectively, to donate money directly to charitable causes.~\cite{speed2022future}. Users can also specify conditional payments using digital forms of money, that, if met, will release funds~\cite{trotter2020smart}.  An example is the Oxfam: If Then Give App. (Figure 1). It lets users track and donate money to specific causes based on set conditions. For example, a user can choose to donate to earthquake relief if an earthquake of magnitude 3.5 or higher occurs in a specific location. Designers continue to experiment with concepts like smart contracts, blockchains, and algorithms that could encourage positive impact and demonstrate the social and global benefits that technology and money can potentially provide~\cite{murray2023blockchain}.

The emergence of technologies that leverage finance in these ways is encouraging and demonstrates a diversity of possibilities, some of which are not solely goal-oriented towards making more money, but rather look to address social concerns. As design explores the innovative domain of digital money, it raises questions about traditional material forms of money that are well-established in society. Evidence suggests that the introduction of new financial technologies can lead to a decrease in the use of traditional financial services~\cite{tok2022fintech}. Vines et al. (2012) have shown that changes in the way people access money can have real-world implications, particularly if individual relationships with money are not considered. In their study with older adults, they found that the material qualities of cash were important, and that cash was tied to social relationships and support. They argue that financial technologies should be designed in an accessible and inclusive way to allow people of all generations to maintain their values in relation to money~\cite{vines2012eighty}.

It is important to acknowledge that technologists working in the FinTech space are not isolated from broader social discourse. As Blumenstock (2018) observes, FinTech is often narrowly conceptualised as a means of facilitating financial transactions, with insufficient attention paid to the social dimensions these technologies provide~\cite{blumenstock2018design}. A phone, for example, has many social capabilities in addition to allowing for money transfers. It can support communication and help nurture and define the quality of social relationships through contextual exchanges of money~\cite{blumenstock2018design}. Monetary transactions, such as sending money to charity or giving a gift, are embedded in social relationships.

Money has always been a social technology~\cite{ingham2020money}, and the emergence of financial technology does not change this fact. FinTech has the potential to change how we view money in society, how we access money in society, and the impact on those who may not have access to digital technology or whose access to money is impeded by its requirements. While FinTech has the potential to positively impact our world, careful consideration is required to ensure that it is used to ameliorate inequalities rather than exacerbate them.
Designers play an important role in shaping the evolving landscape of money by creating technologies that promote financial inclusion, support social relationships and values, and address social concerns such as inequality and exploitation. At the same time, designs must also support the social practices that are relied upon by many, even if the mechanisms required to achieve that support fall outside the scope of FinTech.

\subsection{The Emergence of Central Bank Digital Currencies and values.}

Not all financial technologies are created equal. Schindler (2017) created a table to evaluate the financial impact of different Fintech innovations~\cite{schindler2017fintech}:

 \begin{itemize}

    \item \textit{Surface innovations}, which do not change the fundamental nature of the product or service but perhaps introduce or change a superficial element,

    \item \textit{Genuine innovations}, which change the fundamental nature of a product and thus introduce genuinely new products or services, and

    \item \textit{Foundational innovations}, which introduce significant changes to the infrastructure and underpinnings of the financial system.
\end{itemize}

Not many financial technologies can be classed as foundational innovations, but CBDC falls within this category. CBDC might not only have major implications for the practice of finance, but equally might have significant social, political, and economic impacts. Its functions are different from that of current digital methods of exchange. Like cash, CBDC would be issued by the central bank. Reminiscent of how tactile cash is printed, governed, and distributed to populations, CBDC would be distributed in a similar manner, but in digital form, potentially with the support of third parties. In the UK, this digital ‘coin’ could be known as `digital sterling' or `digital pound'. Due to its potential benefits and impact, CBDC has garnered significant attention and popular interest in recent years.

Countries such as Sweden are on track to becoming a cashless society, as a result of many social and technological factors~\cite{arvidsson2019building,arvidsson2018future}. Digital money, such as CBDC, is poised to shape a new financial landscape, although the current landscape is already undergoing changes. The UK has begun to investigate the concept of a CBDC through a joint effort by the Bank of England and HM Treasury resulting in a series of consultation papers~\cite{bankofengland2020cbdc,boe2021cbdc,digital_pound}. In this report, CBDC is proposed as a new form of payment, and it is assumed that it will not replace cash.  However, although the assumption is welcoming, it is not clear that the Bank of England alone can preserve the role of cash by decree.  Cash is already in decline, with or without CBDC.~\cite{UKDigitalPound2023}. This reduction may have been accelerated by the COVID-19 pandemic~\cite{UKFinance2021}, although it is ostensibly part of a longer cash reduction trend over the past decade or longer~\cite{arvidsson2018future}. Nevertheless, public discussions on CBDC in the UK have not yet seriously evaluated a scenario wherein CBDC amplifies the trend of cash reduction. This is important to consider, because the choice to use cash can embed social, spiritual, family, and individual values~\cite{halloluwa2018value}. As the UK considers the potential benefits of ushering in a CBDC, it is already the case that some forms of money are being displaced, and the introduction of new choices might exacerbate that trend. 

Physical cash remains an essential part of daily life for many people in the UK~\cite{BoE2021}. However, the infrastructures that support access to cash are diminishing~\cite{access_to_cash_review_2019}. Current government policies~\cite{govuk_new_powers_to_protect_access_to_cash} have been found to not adequately address the need to ensure general acceptance of cash and sufficiently convenient ways by which individuals can obtain cash.  The decline in cash usage and the lack of measures to protect it underscore the need to understand how the realisation of community values may change in a world with CBDC.  Ignoring the potential impact could result in the creation of a CBDC that does not accommodate as many social concerns or values as possible.

Accounting for people's current values as the materials of money change is particularly important because, as Swierstra (2013) posits, introducing technological advancements can change our perceptions of morals and values,
with the potential for both negative and positive consequences. The moral outcomes that arise from introducing new technology depend on whether diverse stakeholders are considered and have input in its development.~\cite{swierstra2013nanotechnology}. 
This concept prompts us to consider the individuals who could be impacted by CBDC, including their current values and concerns, as well as whether their perceptions, values, and morals are representational in a CBDC landscape. Unfortunately, the CBDC discussion struggles to break from an economic view of money as a store of value~\cite{bankofengland2020cbdc} or its potential to hold transactional value~\cite{halloluwa2018value}. Undoubtedly, economic value is important to consider, although arguably, the values individuals hold are also relevant~\cite{swierstra2013nanotechnology}, especially when CBDC technologies could have an impact on many retail users of money.

Public discussion of CBDC generally assume that it will be developed from the top down, and the discussions about its design therefore concentrate on its technical specifications, for example, as put by Agur (2020, P, 63), ``Would the CBDC resemble deposits by coming in the form of an account at the central bank, or would it come closer to cash, materialising as a digital token?''~\cite{agur2022designing}
These questions are concerned with how a CBDC will come to be. Although they often do not address the end user of a CBDC, technical questions serve to highlight which stakeholders might or might not be impacted by a CBDC. For example, there are two emergent contexts in which CBDC might be deployed: retail and wholesale. A retail CBDC would be used by people in shops and exchanged with commercial banks, while a wholesale CBDC is traded among banking institutions and is not used by individuals and most businesses. Wholesale CBDC systems already exist, although they do not use Decentralised Ledger Technology (DLT)~\cite{panetta2022demystifying}, which is a secure and transparent way to keep track of transactions. Both wholesale and retail CBDC could come to fruition, althoug designing a retail CBDC is more complex because it involves many stakeholders and considerations~\cite{panetta2022demystifying}, including questions related to human rights such as ownership and privacy.  Therefore, it is specifically crucial for the discourse on retail CBDC to involve all relevant stakeholders, including end users of money, and their values, whereas wholesale uses of CBDC might not require the same degree of consideration.

The design of CBDC and how a person interacts with it will be important to ensure accessibility and meet the needs of its potential users, whether it is wholesale, retail or both. However, being digital makes access to CBDC a challenge from the start. Previous studies in human-computer interaction have shown that some current financial technologies may not meet the requirements of people with certain conditions~\cite{dai2023cognitive}. Therefore, unless digital inequalities are addressed~\cite{gopane2019enquiry}, when a CBDC becomes deployed, pre-existing exclusionary gaps will inherently restrict its access and usability to a certain set of people. For this reason, researchers have argued that the architectural and user aspects of CBDC require understanding~\cite{barzykinclusive}, although work in this area is still scarce. Consequently, it might be necessary to apply design perspectives and methods, such as value-led and inclusive design. These methods focus on the values and needs of end-users and aim to make design outputs more accessible and reduce social barriers for diverse populations~\cite{iversen2012values,clarkson2013inclusive}. The success or failure of a CBDC will depend on whether it aligns with the values of its users, as design outputs only become ‘grounded’ in everyday practices if they reflect the users' current and emerging values~\cite{iversen2012values}. Valuing a diversity of users will require taking into account a diversity of conditions~\cite{erazo2015design}, to ensure that everyone can access digital and physical infrastructures in a world with CBDC.

The uncertainty surrounding CBDC remains substantial.  The novel digital coin raises many questions, many of which remain financial~\cite{auer2021cbdc}, leaving many alternative questions unanswered or unconsidered. Nevertheless, designers argue for embracing uncertainty~\cite{soden2020embracing}, as it can lead to periods of inquiry~\cite{vandePoel2022} and inclusive design outputs~\cite{bowler2022exploring}. Attending to uncertainty can create awareness around processes of change~\cite{pink2018uncertainty} and doing so is becoming a welcome quality of the design process~\cite{akama2015design+}. As financial policy around CBDC moves quicker than the speed it takes to do CBDC research~\cite{auer2021cbdc}, addressing the uncertainties present in CBDC discourse presents a challenge.  Despite this, the purpose of embracing uncertainty is not to answer the uncertainty or to create certainty, but to reveal the unknown~\cite{bowler2022exploring}. To achieve this, designers might need to take a more flexible approach to CBDC, using both trans-disciplinary and interdisciplinary methods, and incorporating a diversity of stakeholders. This could help us understand what we know and what we don’t know while accepting that some things might remain unknown~\cite{chua2002known}.

There is still much we have yet to discover. However, we know, as previously discussed, that as CBDC pushes into society, digital inequality remains a pressing issue, and certain preconditions to accessibility are being overlooked.  Money is deeply intertwined with all areas of society, and people hold diverse values in it beyond its numerical worth. Established forms of money are steadily being displaced and technologies can have significance for values and morals.  Such factors make the design of novel money a grand-scale challenge that requires attentiveness and criticality. Without such an approach, there is a risk of detrimentally impacting millions of people. One way to address this challenge might be to consider money as a common design material: a material that goes beyond its physical properties. Even as money moves into the digital realm, it is crucial to maintain its status as the most common design material. This will ensure that a diversity of manifestations of money provide value, multiple avenues of access, and a financially equitable social landscape. It is also important to hold new forms of money accountable for their material and immaterial properties.

\subsection{Stakeholders and Methods}

To better understand the role of design in the CBDC landscape, we reached out to stakeholders involved with CBDC in various capacities. These stakeholders included representatives of UK public bodies, Web 3.0 specialists, executives with expertise in commercial banking, and digital payment industry professionals. Despite the small sample size of four participants, their knowledge and expertise provided valuable insights, which encourage designers to explore current established forms of money and a future world with CBDC.

We conducted remote semi-structured interviews and captured data via audio recording and transcription. The data were then anonymised, collated, and uploaded into Nvivo 14 for analysis using a constructivist grounded theory approach~\cite{charmaz2006constructing}. According to Charmaz (2014, p.131), this approach can reveal ``the connections between micro and macro levels of analysis'' and link ``the subjective and the social''~\cite{charmaz2006constructing}.  This makes it an appropriate method for extracting how an individual’s micro subjective insights reflect upon macro social knowledge, values, and global considerations with the emergence of CBDC. The data analysis process involved conducting a line-by-line assessment of each transcript to formulate initial codes. We then refined these codes through an axial coding process and cross-referenced their relationships with codes from other participant transcripts. Finally, we used selective coding to identify core categories and present emergent theories in our findings section.
We have identified themes that can aid designers in understanding CBDC from various angles, including its potential for value creation and inclusivity as well as its cautionary tales. The following section reports on these findings.

\section{Findings}

\subsection{Reforming Values through the Design of Central Bank Digital Currency}

Our research suggests that CBDC presents a unique opportunity to reform public values through its design. As one participant stated, “We’re not going to get better money with CBDC unless we design it to be a better one” (Participant 3). This is not solely due to the technical properties of CBDC, but rather the chance it provides society to have money represent more of its values than it currently does.

One participant referred to current monetary infrastructures as ``haphazard and a bit disorganized,'' leading to issues like ``oppression and exclusion'' (Participant 1). CBDC was seen as an opportune moment to reinvent the values of money. As that participant put it, ``Let’s think about the role that Digital currency can play in creating the sort of society that we actually want'' (Participant 1). Yet, time to add such values might be limited, as the introduction of CBDC into society is just a matter of time. As another participant stated, ``The stone will be thrown. We don’t know where it’s gonna land and we don’t know how deep it’s gonna go. But we do know there’ll be ripples from it'' (Participant 4).

The unknown ripples from CBDC mean that its trajectory and value will only become clearer as it comes into being. Yet, some predictions might be attributed to its positive value impact. For instance, CBDC could track commercial practices from production to consumption using smart technologies and influence behaviour for the better. As one participant explained, ``Digitalisation of trade [...] to increase the right sort of growth and enable greater demonstration by companies and traders of their ESG compliance and good ethical and carbon neutral behaviour'' (Participant 1). This was seen as an opportunity to offer companies VAT relief for services that are not carbon-heavy. Essentially, ``you could affect people’s behaviour for the better'' (Participant 1).

The potential for CBDC to influence behaviour, while potentially positive, also presents concerns. As one participant stated, ``Some of the folks feel that we have already given away so much of our privacy that it wouldn’t be a problem to have a central entity like the Central Bank, [...] you know, take a peek on, you know how many pints I bought tonight. [...] Do we want to have that at the core of our financial system?'' (Participant 3) CBDC can do wonderful things for concepts like sustainability, but it can also penalise people if their practices do not align with the governed values present in the CBDC design, potentially leading to harmful practices. As one participant expressed concern, “I’m concerned about privacy and protections [...] we need to prevent abuses from happening'' (Participant 2).

The nature of CBDC can allow for more transparency of people’s actions, dependent on the forms of its deployment and governance, which will determine the negative or positive effects of CBDC. This leads some to suggest ``There needs to be that full debate about how much privacy and transparency we need'' (Participant 1). An option to address this was seen to be designing around flexibility. As one participant suggested, ``I think you would design in enough flexibility that you can have varying degrees of privacy and transparency'' (Participant 1). Those in charge of CBDC designs were deemed to need to offer honest transparency about their knowledge. As one participant stated, ``There will be privacy to it but it creates such a grey area and that’s the problem [...] so much vagueness at the moment and that’s fine as long as we are fully confident and open and honest that it is really, really vague and we don’t have the answers'' (Participant 4).

The uptake of a CBDC might depend on the values people are seeking in a CBDC. As put by one participant, ``The best way to tackle innovation is to build something that is dramatically better than what exists and people will just move because it doesn’t make sense to be in the old world anymore'' (Participant 3). If a CBDC can meet this standard, people might naturally migrate to it. The value from CBDC will depend on how its design and implementation are approached. As one participant explained, ``I think the real risk is that we treat it just as being a marginal increase to the efficiency of our rather complex payments and settlement systems and that we miss the real opportunity [...] its role as a policy delivery [...] that’s both social and economic as well as sort of government policy'' (Participant 1).

For participants, CBDC was not just an aesthetic change to money in digital form but an opportunity to reinvent its social representations that could drive new notions of inclusive, ethical and global values. Nevertheless, it’s evident CBDC is a fine balance between what and whose values are being re-imagined and equally whose are not.

\subsection{The Consideration of Time in the Design and Deployment of CBDC}

The speed at which a CBDC is deployed, discussed, and regulated can impact the values reflected in its design. ``I feel like it's been a very rapid acceleration to where we are. It's been very focused on the kind of the consultation [Bank of England] paper reading'' (Participant 4).  This echoed a reflection of another participant: ``The debate about CBDC has got too specific too quickly in my view and is currently mouldering in the regulatory space, and it really shouldn’t be there at all. This should be a civil society debate'' (participant 1). Rushing the process may limit the involvement of society in the conversation.

Aiming to be first or reacting to the progress of others could also result in a CBDC that does not fully represent the values of its users, ``half-baked because you haven’t given it enough thought'' (Participant 3). Having time to think through a value-centred approach and ensuring that the CBDC represents ethical, sustainable, and future-oriented values could lead to greater uptake of a digital coin that ``people will want to use to demonstrate [is] ethical, [is] sustainable, and [is] their future'' (Participant 1). Pace becomes an important consideration in the design of a Central Bank Digital Currency. Taking the time to ensure that values and needs are built into the design, if these values and needs are not incorporated, then the uptake of the coin may be jeopardised.

One current expectation of money is that the rate at which it moves around can be slowed and ``limited by the movement of physical paper currency,'' according to Participant 2. This can be positive in slowing down other central banks from doing a run on another central bank’s currency.  Participant 2 explained that central banks often hold each other’s currencies. However, if the task of directly holding a foreign currency becomes unencumbered by the limitations of physical assets, then moving money can become instantaneous, posing a problem. ``So you then have the situation where a central bank is no longer able to prevent a run on its currency because other central banks can just send them off to other partners at will without another central bank being able to stop it,'' said Participant 2.

Bank runs are a popular position to conceptualise why friction might be required. Participant 3 asked, ``What if you have a bank run? Does a tokenised deposit make it worse? Does the CBDC make it worse? What type of new friction do we have to bake into the system if everything now runs faster with less friction? Is it that we’re going to bake in cool-down periods?''  Time became a useful asset to prevent negative impacts. ``Do we need to introduce some kind of delayed mechanism for settlement in Central bank digital currencies in order to allow netting of transactions across different boundaries in foreign currencies?'' asked Participant 2.

The quickness of a CBDC was also seen as a positive. ``Frictionless experience for somebody who’s wanting to trade across a border, especially for the small-to-medium enterprise sector, who so far are showing little appetite for doing it,'' said Participant 1, suggesting that the speed of CBDC and the frictions designed into it are a matter of context.

\subsection{Echoing Present Values in the Design of CBDC}


The present values associated with money are vast.  Money carries specific personal or social values with it as it circulates, meaning that questions such as the one asked by Participant 4 become important in CBDC discourse. ``How do we get the value of the automation from digital currency without losing the Community sense that you get from cash, you know, how it interacts with the community?'' (Participant 4). Cash became a signpost to echo present values. ``Immediately and anonymously, I can get cash out of the bank [...] I can give somebody a £10 note and my debt is immediately anonymously and Frictionlessly settled.'' (Participant 1). Cash offered privacy, immediacy and other affordances; however, a CBDC emergence did not necessarily equate to the removal of cash. ``I think there's going to be less need for cash, but there's still going to be a need for it'' (Participant 2). Though the use case for cash would still be needed, its access could become challenging, as the continued statement by Participant 2 implies: "I think the supporting infrastructure for that, is going to dwindle [...] probably will start to be increasing fees for the use of ATMs as there has already been in other countries.''

But cash offers other ``quality differences'' (participant 2) that CBDC ought to consider. ``Cash, for example, is more efficient to budget on, so it's more important to have that during times of financial crises. So, I'm hoping [...] a digital pound would serve those kinds of financial budgeting capabilities and enable people to better control their money'' (Participant 4).  Cash has many values connected to it that are beneficial and hence desirable features for a CBDC design to try to replicate. Generationally people have come to form an idea of their money. ``You know, paper money because they're still generations that are used to that. There is a cultural, societal aspect to it that values that right and the shocker to me when I arrived in the UK was like Places that only accepted contactless payments. In Brazil,  if you're a business and you don't accept cash, bank notes it's a federal crime.'' (Participant 3). People's present values in cash might rely on policy and law to protect the values intrinsic to current forms of money.

Familiarity with how people interact with money is an important prerequisite for the design of CBDC. Distribution of CBDC from a central bank meant ``You now have a liability against the central entity. That's a sovereign agency. Um, which is a relationship. Most citizens are not used to'' (Participant 3). People do not bank with the central bank, but rather commercial banks, which provide choice, and flexibility and are detached from sovereign issuers.  It also exposes everyday people to risk: If the transaction mechanism for a CBDC were centrally controlled, then money would come under a new form of control. ``I click a button and I just suspend everything from the system, which is like a very like last resort type of measure'' (Participant 3).  CBDC can offer central authorities the ability to freeze or stop transactions, which then introduces questions of trust and ownership. ``But then it also tells the people that their money isn't theirs. And that is something that money in the physical form has told people for centuries now. And if we change that, we'll probably be infringing the very nature of money'' (Participant 3). The nature of money has been built on trust that the money you have is real. ``The coins you get, you have to trust that they're made out of things. The notes you receive and change from the local shop, You have to trust that they're not counterfeit, right? [...] You just have to trust somebody sometime'' (Participant 2). This also formulates into ownership, people need in some form to feel they have ownership over the money in their possession. CBDC, however, ``changes how we need to think about the design of money and it forces us to think about who owns the money that we are supposedly owning. Is it us? Is it not us?  And if it's not us, how to ensure that the money that I have in my possession still remains in my possession and is accepted universally [...] These are all big questions and I don't think. Most of, and this is from the readings that I've from my BIS readings. I don't think we're having those conversations globally'' (Participant 3).

Equally, there is a concern that, although CBDC presents an opportunity to have impact on the way money is used, the opportunity might be missed.  As put by Participant 1, ``It would be a real shame if the only outcome of CBDC was it as a wholesale one, and it made [commerical banks] 1 per cent more profit. That would be a real shame.'' The present situation shines light on the features designers might want to incorporate into a CBDC, or try to change for the better.

\subsection{Promoting Inclusion in the Deployment of CBDC}

The design of the CBDC system that emerges is important in determining the inclusionary properties afforded or not afforded by CBDC. Participant 3 explains that ``If it’s wholesale [CBDC], you’re talking to people that are in the top 0.1 per cent of capitalism [...] I don’t think we need to involve, you know, John Smith from down the street to discuss wholesale CBDC.'' Wholesale and retail CBDC were seen to require inputs from different users, with different requirements.  Moreover, the extent to which inclusion applies is dependent on the context of its deployment.

``Imagine you’re a commercial bank, right? You have your demographic that, you know [...] You have a smaller population than the whole central bank would have [...] You know your demographics that you can tap into at your scale and you can design for them a tokenised deposit experience that makes sense to those customers'' (Participant 3).  For commercial banks, a focus on users was seen to be an agreed consensus. ``I think customer-centricity is vital. You know, as early as possible [...] I think the user focus is: no we don’t [focus on users because] we are very early on, knowing around what the CBDC does'' (Participant 4).  Grasping an idea of what the CBDC does and what the government, market, and economy decide to do with CBDC directly influences which users become relevant to its inclusive potential.

Equally important is considering which users will get left behind with the manner in whic CBDC is deployed and updated.  Being digital, CBDC will require technology updates, security patches, and so on.  User focus becomes important here, as pointed out by Participant 2: ``How do we update it? [...] Do you get people to convert from that version to another version? Do you fork it into a version that uses a new algorithm? Then what happens to the users that get left behind or don’t know about it?''  It is clear that the specific design of the CBDC infrastructure, its versions, and how it is deployed all become avenues to question what kind of user focus is required, and where this applies within the wider notions of CBDC discourse.

With regard to how CBDC is discussed, inclusion depends upon simple, jargon-free language: ``I think is really important for the progress of this, is the inclusiveness of the definitions of the language. I think that is crucial. That is vital. That you need to have everyone agree that we use the same language and it's inclusive. It's not jargony'' (Participant 4).  The current CBDC discussion comes with a complexity of terms and technological knowledge; simplifying and aligning terms and meanings could offer those not able to join such discussions the opportunity to do so. Simplifying the description of systems around CBDC also lends itself to a degree of user focus that currently is not present. CBDC could end up functioning in ``Web 3.0'', wherein the user experience is deemed ``ridiculous'' (Participant 3). People today value the user interfaces to which they have become accustomed with online banking, meaning ``my thinking is whoever fixes [user interfaces] in Web 3.0 wins'' (Participant 3). Equally, education was seen to be a pivotal part of CBDC deployment and inclusion: ``You hope that it isn't going to benefit a minority of  users. And I think that the key to doing that is the education piece. You can't just, you know, you can't just expect it to be lay, or self-assuming because it's there, people will use it'' Participant 4).  It was deemed that this would be the role of commercial banks to provide and not that of the central bank. Ensuring inclusive access to both discussion, education and user-friendliness will require effort in all areas, from the design of CBDC user interfaces to the language, to education approaches used in and around CBDC discourse.

\section{Closing remarks}

This paper aimed to initiate a discussion among designers about their role in influencing or preserving the values of money, both in its traditional forms and in its emerging form of Central Bank Digital Currency.  However, this study has some limitations.  One is the small sample size of only four participants. As a result, the findings should be viewed as suggestions for further exploration rather than conclusive evidence. Another limitation is that all participants had some familiarity with CBDC.  Finally, further research with a larger and more diverse group of participants, including those with and without prior knowledge of CBDC, could provide more insight and can broaden our understanding of how it is perceived and reveal new concepts of value and the materiality of money.

Nevertheless, our findings reveal the challenges and opportunities that CBDC poses for design. They underscore the potential of design to shape this evolving landscape. We have addressed certain areas where future work could be taken by designers:

\begin{itemize}

\item\textit{CBDC presents an opportunity to shape community values.} However, this must be approached with caution. Our findings suggest that people may have differing ideas about what constitutes value. If one group has control over creating or upholding value, others may be subject to oppressive experiences. Design methodologies, such as participatory and co-design, could be utilised to bring stakeholders together and facilitate discussions between a diverse range of people, including those driving CBDC discourse and those who have yet to encounter the concept. This could create a dialogue that merges the opportunity presented by CBDC to shape positive values while maintaining the sovereignty, individuality, respect, and integrity of individuals within a society. As Participant 3 mentioned, it is also important to have these conversations on a global scale.

\item\textit{The language used when describing CBDC and how it works should be simpler.}  Our study found that bringing together a diverse range of people to discuss CBDC may present challenges due to the topic’s complexity. As Participant 4 pointed out, it is important to start building inclusive language and terminology around the concept of CBDC. Design may play a role in shaping and helping to create the lexicon associated with CBDC. This could involve designing materials, toolkits, or prototypes that enable people to build a language for discussing CBDC or create a CBDC language that is understandable to many. This may need to be cross-generational and take into account a diversity of people and approaches. Additionally, design may end up supporting education about CBDC, along with commercial banks, which will be needed in order for people to engage with CBDC and use it well.

\item\textit{The uncertainty surrounding the nature of CBDC is significant.}  Whether wholesale or retail, whether token-based or account-based, makes designing for CBDC a challenging task. Yet, as previous designers and researchers have shown~\cite{pink2018uncertainty,akama2015design+,soden2020embracing}, uncertainty can be a useful phenomenon to have within the design process. Design methods, such as speculative design or design fiction, could be utilised to envision uncertain scenarios and futures where each of these versions comes to light. This would allow for the interrogation of each version to understand its positives and negatives while remaining in the present, before deployment of a specific new form of money has taken place. This could help circumvent any potential negative consequences that were unimaginable and only come to light once a person starts using one of these methods of money.

\item\textit{CBDC presents an opportunity to sustain and support existing values.}  The values associated with money will play a crucial role in shaping the design of CBDC. Factors such as trust, policy, familiarity, and experience will be key considerations for CBDC design. This will be a fine balance for CBDC, as removing  existing values could be detrimental to the more than numeric value people place within the discourses of money in maintaining their financial routines~\cite{vines2012eighty} and having the infrastructure to access their tangible assets~\cite{tok2022fintech}.  Future work might involve designers finding ways to better understand existing values surrounding discourse about money, ensuring that CBDC reflects those values through its design. If CBDC is unable to reflect those values, designers might want to explore new ways to design for existing forms of money, such as cash, in order to maintain its tangible existence within society, as many people still rely on it. Designers might also want to consider the tangible aspects of money, to determine whether and how certain precautions and frictions should be incorporated into the design of CBDC, since the tangible properties of cash are not entirely reflected by a CBDC unless a physical concept of a CBDC is created.

\item\textit{The temporal properties of CBDC are important.}  Designers may want to consider Temporal Design~\cite{pschetz2018temporal} as a way to explore the various temporal properties associated with CBDC. For example, Participant 1 mentioned that work on CBDC has in some ways progressed too quickly.  Since CBDC is not a tangible material, its temporal qualities are very different to that of cash. It has the potential to transfer money at unprecedented speeds, and whether it is necessary to explicitly introduce friction becomes an important design consideration. Exploring the temporalities of CBDC could provide interesting opportunities for future work and could illuminate ways to address the challenge of political decisions about CBDC outpacing the research~\cite{auer2021cbdc}.

\item\textit{CBDC offers a chance to focus on inclusivity.} Although some efforts have been made to understand inclusion and exclusion to a great extent, these have largely been theoretical and have not incorporated user-driven perspectives~\cite{barzykinclusive}. A theme that emerged from our findings is the connection between value and inclusion and the need for more user-centric CBDC design. There are many challenges to overcome in this regard, including the speed at which political decisions are evolving faster than research can keep up,~\cite{auer2021cbdc}, the diversification of values around money, the use of jargon, education, and issues identified in the literature, such as digital exclusion~\cite{lloyds2022consumer},~\cite{dai2023cognitive,gopane2019enquiry} and its relationship to financial exclusion. CBDC is more than just a new form of money: its potential could be significantly positive, as reported by Participant 1.  However, without user-led design, its potential to be detrimental is equally significant. Future work that takes into account our previous considerations have an opportunity to promote inclusion. Moreover, we can identify the types of users who should have a say in the design of CBDC, with the potential to influence CBDC discourse to become a more inclusive emergent concept. Without design, CBDC might become simply another avenue ``for money to make more money'' (Participant 1).

\end{itemize}

In conclusion, money has taken many forms throughout history, including non-tangible representations. The rise of Central Bank Digital Currency (CBDC) will transform our understanding of digital currency. Money is connected to values: social, cultural, religious, moral, and more. However, new forms of money can lead to the demise of other forms of money and their associated values.  Design can bridge the gaps in inclusivity that CBDC might otherwise create and shape new positive and inclusive values. Now is an unprecedented moment to re-envision the meaning of money. We may need time to consider how money came to be and begin ``reverse engineering'' (Participant 3) before we can understand what values CBDC does and does not represent, and more importantly, \textit{whose} values are represented by CBDC. Design in the space of CBDC could inform a more equitable form of money while actively supporting current and established notions of money. As one participant stated, ``the stone will be thrown; we do know there’ll be ripples from it'' (Participant 2).  Design has a moment to shape the ripples made by Central Bank Digital Currency and help establish a more equitable and inclusive social, political and global foundation associated with money. It is important to consider who has a stake in this most common design material, money, as digital sovereign currency emerges on the world stage.





\raggedright
\bibliographystyle{plain}
\bibliography{bib.bib}

\end{document}